# Does Global Neuronal Workspace Theory Explain Phenomenal Consciousness? The Motivated Emotional Mind Challenge

**Wiesław L. Galus & Janusz A. Starzyk**

## Abstract

Global Neuronal Workspace Theory (GNWT) is one of the most extensively developed empirical research programmes on access consciousness. By contrast, the Motivated Emotional Mind (MEM) model proposes an embodied, semi-hierarchical associative memory in which representational selection, action, recurrent reconstruction of modality-specific fields, interoception, and valence form a single functional cycle. This article assesses whether MEM mechanisms can reproduce the functions explained by GNWT and whether they yield additional hypotheses concerning phenomenal quality. We first reconstruct the origins, mechanistic core, strengths, and limitations of GNWT, and then present the origins of MEM, the architecture of semblions, and the FFS–RP cycle. In the comparative analysis, we distinguish functional similarity from mechanistic identity and empirical equivalence. We examine 25 publications: twelve canonical studies of GNWT or critiques thereof, and thirteen studies of perception, interoception, and affect. For each study, we juxtapose the GNWT interpretation with a possible MEM reinterpretation. The findings indicate that most classical effects—including threshold dynamics, maintenance, recurrence, long-range connectivity, and flexible use of information—are compatible with both theories. This compatibility weakens the claim that the GNWT interpretation is exclusive, but does not yet provide theory-specific support for MEM. MEM advances a broader hypothesis integrating access, modality-specific reconstruction, bodily state, affect, and action; its potential explanatory advantage nevertheless remains a hypothesis requiring preregistered quantitative tests. We propose a series of interventions designed to dissociate sensory re-entry, global availability, report, and interoceptive valuation.

**Keywords:** Motivated Emotional Mind; Global Neuronal Workspace Theory; access consciousness; phenomenal consciousness; re-entry; interoception; semblion

## 1. Introduction: The Comparative Question and Evidential Asymmetry

The purpose of this article is to assess whether the Motivated Emotional Mind (MEM) model can reproduce the functions that Global Neuronal Workspace Theory (GNWT) attributes to access consciousness, and whether it yields additional, empirically discriminable hypotheses concerning phenomenal quality. The motivation for this work is the MEM model developed over the last decade (Galus, 2015a, 2015b, 2025a, 2025b, 2025c, 2026; Galus & Starzyk, 2020, 2026, Starzyk & Galus 2026), which aspires to integrate existing theories of consciousness based on well-established neurological and biophysical theories and hypotheses, among them hierarchical sensory processing, recurrent processing, synaptic plasticity, distributed and sparse coding, interoception, and the influence of affect on learning and action. A model that claims such an integrative role must be confronted with the most mature contemporary research programmes. GNWT, as the most extensively developed empirical account of conscious access, is a natural and demanding point of comparison.

The MEM architecture contains candidate mechanisms for all the principal functions that GNWT attributes to access consciousness: selection among competing contents, threshold amplification, temporary maintenance, access to memory, report, and control of action. At the same time, MEM goes further by attempting to relate these functions to the modality-specific content of percepts, qualia, mental images, and affective states. Breadth of scope alone, however, does not establish empirical superiority. GNWT has developed over approximately four decades and has inspired an extensive psychophysical, electrophysiological, neuroimaging, clinical, and computational research programme. MEM is a younger programme, and some of its distinctive hypotheses have not yet been investigated in studies designed specifically to falsify them.

Three questions must therefore be distinguished. The first concerns scope: can MEM explain, at the functional level, at least the access-related phenomena explained by GNWT? The second concerns mechanism: do the semblions, winner-take-all (WTA) competition, feedforward sweep (FFS), recurrent processes (re-entry; RP), secondary perception, and allostatic valuation proposed by MEM constitute an explanation distinct from neuronal ignition and global broadcasting? The third concerns evidence: are there findings that discriminate between these mechanisms, or only findings compatible with both accounts?

A direct MEM research programme should meet the standards of rigour achieved in GNWT research: preregistration, quantitative temporal and spatial predictions, causal interventions, no-report paradigms, and out-of-sample model comparison. Until such a body of evidence has been assembled, an indirect analysis remains useful. It consists in re-examining canonical experiments designed to test GNWT and determining whether their results also follow from the MEM architecture. Because the core GNWT literature has focused more on access, threshold dynamics, and report than on the genesis of modality-specific content, we add a separate set of studies on perception, interoception, and affect.

We adopt a cautious standard of inference. The reputation of the laboratories involved and the replicability of the basic effects do not guarantee the correctness of any single theoretical interpretation. If the same result is predicted by two models, it increases the plausibility of the shared class of mechanisms but does not adjudicate between the theories. Reinterpreting GNWT data in terms of MEM may demonstrate theoretical underdetermination and the feasibility of an alternative explanation; it cannot replace a theory-specific test of MEM. Our final claim therefore takes the form of a hypothesis that MEM has greater explanatory potential, rather than a declaration of demonstrated superiority.

The remainder of the article is organized as follows. Section 2 reconstructs the origins, empirical basis, strengths, and unresolved problems of GWT and GNWT. Section 3 presents the origins of MEM, its architecture and morphology, the common mechanistic scheme of the FFS–RP cycle, the principal mental processes it addresses, and the scope of its unification and its limitations. Section 4 analyses the mechanistic relationship between the two theories: it considers the transition from the global workspace to an embodied network of semblions, the treatment of perception, qualia, and motivation, secondary perception and report, the scope of complete and incomplete functional substitution, and the distinction between shared, orthogonal, and genuinely competing claims. Section 5 describes the method of comparing the empirical evidence: the corpus of 25 publications, sixteen theory-neutral criteria in four dimensions, the rating scale, and the coefficients of explanatory scope (M), evidential consistency (E), and conservative fit (K). Sections 6 and 7 examine the corpus itself, that is, twelve canonical studies of GNWT or critiques thereof, and thirteen studies of perception, interoception, and affect, and juxtapose the GNWT interpretation of each with a possible MEM reinterpretation. Section 8 reports the structured results of the comparison, and Section 9 discusses their significance for access consciousness, qualia and phenomenal consciousness, and interoception, allostasis, and motivation, as well as the meaning of MEM's present 'advantage'. Section 10 proposes an experimental programme to distinguish MEM from GNWT and specifies criteria for falsifying MEM. Section 11 presents the conclusions. Appendix A contains the publication rating matrices.

## 2. Global Workspace Theory: Origins, Development, Strengths, and Limitations

Global Workspace Theory (GWT) and its neuronal development, Global Neuronal Workspace Theory (GNWT), are among the most influential research programmes on consciousness. Their central claim is that numerous specialized brain processes operate in parallel and largely unconsciously, whereas a content becomes consciously accessible when it is selected, maintained, and made broadly available to systems responsible for memory, decision-making, report, and action control. In GNWT, this transition is associated with the nonlinear 'ignition' of a distributed network and long-range recurrent communication (Dehaene, Kerszberg, & Changeux, 1998; Dehaene & Naccache, 2001; Mashour, Roelfsema, Changeux, & Dehaene, 2020).

GWT and GNWT are not interchangeable concepts. GWT is primarily a functional theory of the global availability of information, whereas GNWT links this principle to neuroanatomical, dynamical, and synaptic hypotheses and, more recently, to connectomic, receptor-level, developmental, and valence-related hypotheses. The theory most precisely characterizes access consciousness: the conditions under which a representation can flexibly influence multiple systems. It remains more controversial whether global availability is sufficient to explain the phenomenal quality of experience.

Historically, the first concise presentation of GWT was Baars's 1983 chapter 'Conscious Contents Provide the Nervous System with Coherent, Global Information.' Baars proposed that the brain comprises numerous partly autonomous, specialized processors and that the function of conscious contents is to provide them with shared, globally available information. Consciousness was therefore identified not with a single centre, but with a limited functional workspace in which competing representations acquire the capacity for widespread influence (Baars, 1983). A comprehensive account appeared in A Cognitive Theory of Consciousness (Baars, 1988), which developed the concepts of competition, process coalitions, limited capacity, and global broadcasting.

Baars's later publications clarified that the global workspace is neither an anatomical 'theatre' nor a central store. The theatre metaphor was intended to represent the selection of a single dominant content and its availability to a large 'audience' of unconscious systems (Baars, 1997). The conscious access hypothesis characterized consciousness as a mechanism for mobilizing and integrating functions required particularly in novel, non-automatic situations (Baars, 2002, 2005). From its inception, GWT was therefore primarily a theory of the functional role of consciousness rather than a complete theory of its neural substrate or phenomenal quality.

The transition to a neuronal model was prepared by studies of networks that learn sequences and by formal models of cognitive functions (Dehaene, Changeux, & Nadal, 1987; Changeux & Dehaene, 1989), as well as by an early neural global workspace model linking attention to distributed thalamocortical integration (Newman, Baars, & Cho, 1997). In the strict sense, however, the foundational publication for GNWT was the article by Dehaene, Kerszberg, and Changeux (1998). The authors contrasted local processors with a distributed population of neurons characterized by long axons and strong reciprocal connections. Access to this network was proposed to enable the maintenance of a selected representation and its transmission to multiple systems, particularly in tasks requiring effort, flexible coordination of operations, and departure from automatic routines.

Dehaene and Naccache (2001) transformed this proposal into a programme for cognitive neuroscience. Masking studies showed that unconscious stimuli can undergo semantic processing, but their effects remain weaker, shorter-lived, and less widespread (Dehaene et al., 2001). Network models linked subjective report to crossing a threshold of recurrent amplification (Dehaene, Sergent, & Changeux, 2003), and related spontaneous background activity to fluctuations in access to an identical stimulus (Dehaene & Changeux, 2005). They also consolidated the distinction between subliminal, preconscious, and conscious processing: preconscious content can gain access, but does not do so without sufficient attentional amplification (Dehaene, Changeux, Naccache, Sackur, & Sergent, 2006).

## 2.1. Current Status and Empirical Basis of GNWT

In mature GNWT, conscious access is a dynamic state rather than a location. Representations compete for a limited workspace; the winning content crosses a nonlinear threshold, triggers recurrent amplification, and becomes globally usable. In attentional-blink studies, conscious recognition exhibited bifurcation-like dynamics accompanied by late and widespread activity (Sergent & Dehaene, 2004; Sergent, Baillet, & Dehaene, 2005). A nonlinear access threshold was also demonstrated using MEG and EEG (Del Cul, Baillet, & Dehaene, 2007), while intracranial recordings revealed a convergence of sustained potentials, high-frequency activity, phase synchronization, and long-range connectivity (Gaillard et al., 2009).

The synthesis by Dehaene and Changeux (2011) codified the canonical markers of access: late nonlinear amplification, the P3b component, high-frequency activity, and communication between distant regions. The programme was subsequently extended to working memory, cognitive control, anaesthesia, and disorders of consciousness. The global workspace was interpreted as a mechanism of compression and flexible information routing (Dehaene, Charles, King, & Marti, 2014), and multidimensional EEG markers were applied clinically in patients unable to communicate (Sitt et al., 2014). Mashour and colleagues (2020), however, emphasized that the theory's core is neither a single marker nor a fixed frontoparietal map, but a nonlinear, recurrent transition to global availability.

Recent work has shifted GNWT from the general metaphor of broadcasting towards formal, biologically constrained models. The Predictive Global Neuronal Workspace describes access in the language of active inference as competition among high-precision perceptual hypotheses (Whyte & Smith, 2021). Klatzmann and colleagues (2025) constructed a whole-cortex model based on the connectome and AMPA/NMDA receptor gradients, in which connectivity and synaptic properties generate a bifurcation between rapidly decaying activity and widespread, sustained ignition. This result increases the biological plausibility of the mechanism, but does not by itself demonstrate that the modelled state is sufficient for phenomenal experience.

The largest contemporary test of the theory was the multicentre adversarial collaboration between GNWT and IIT. The study found coding of conscious content in visual, temporal, and partly frontal regions, together with fronto-visual synchronization, but did not confirm all the predicted patterns of ignition and prefrontal coding (Cogitate Consortium et al., 2025). Proponents of GNWT responded that the study used only clearly visible stimuli and did not test the central conscious–unconscious contrast, and that some of the predictions examined were not part of the theory's core (Naccache et al., 2025). The findings therefore require revision of some strong interpretations, but do not constitute a straightforward refutation of GNWT.

The most recent self-interpretation presents GNWT as a multilevel model encompassing relations among receptors, cells, circuits, the connectome, global state, learning, development, and valence (Changeux & Farisco, 2026). This expansion addresses the charge of excessive functionalism, but simultaneously increases the need to distinguish clearly between the core theory and auxiliary hypotheses. Contemporary reviews and databases confirm GNWT's central position in the debate while indicating that direct tests discriminating among competing theories remain rare (Seth & Bayne, 2022; Yaron, Melloni, Pitts, & Mudrik, 2022; Mudrik et al., 2025).

### 2.2. Strengths and Principal Domain

GNWT is dominant primarily as a theory of conscious access, reportability, and the flexible use of information, rather than as an uncontested, complete theory of phenomenal consciousness. Its strength lies in combining an intuitive architecture with testable predictions concerning threshold behaviour, timing, maintenance, and long-range connectivity. The same conceptual apparatus can be applied in psychophysics, EEG/MEG, fMRI, intracranial recordings, computational modelling, and clinical diagnostics. The theory also acknowledges rich unconscious processing and therefore does not reduce consciousness to the mere activation of a representation.

The programme's enduring appeal also stems from its cumulative character. GNWT generates models, experimental paradigms, and markers that can be revised without abandoning the general principle of global availability. It provides a language linking cognitive and neuronal levels, inspires clinical applications, and supports integration with predictive processing and active inference. Moreover, it is formulated precisely enough to be criticized and tested, although stricter falsification criteria, preregistered predictions, and clearer separation of core from implementation claims are still required (Doerig, Schurger, & Herzog, 2021).

### 2.3. Weaknesses and Unresolved Problems

The most serious limitation is explanatory: global access explains why information can guide report, memory, and action, but does not determine why it is accompanied by a particular 'what-it-is-like' quality. Naccache (2018) defends the claim that sufficiently rich access consciousness can encompass phenomenality. Block (2007, 2011) and Lamme (2006, 2010), by contrast, maintain that experience may be richer than, or precede, access and report, and that local recurrent processing may suffice for phenomenal vision. The controversy remains unresolved because the absence of report is not equivalent to the absence of experience.

A second problem is distinguishing the neural correlate of consciousness proper from its prerequisites and subsequent consequences (Aru, Bachmann, Singer, & Melloni, 2012). P3b and some frontoparietal activity are attenuated in paradigms without immediate report, suggesting an association with decision-making, working memory, or reporting (Pitts, Metzler, & Hillyard, 2014; Tsuchiya, Wilke, Frässle, & Lamme, 2015). It also remains unresolved whether prefrontal cortex is necessary for experience itself or mainly for metacognition and control; evidence supports an important role of posterior cortex in coding content (Boly et al., 2017; Cogitate Consortium et al., 2025).

Ignition, recurrence, and global synchronization are not specific to consciousness; they also occur in models of attention, decision-making, and working memory. GNWT must therefore specify additional conditions that distinguish a constitutive mechanism from a correlate of the global use of information. Explanations of the modality-specific quality of content, feature binding, embodiment, interoception, affect, and the minimal conditions of consciousness in infants, animals, and artificial systems also remain less developed. The multilevel extension of the theory acknowledges some of these problems, but does not yet yield predictions for them as precise as those concerning conscious access.

Despite these problems, GNWT remains a mature, empirically productive, and actively developing programme. Its best-supported domain encompasses selection, threshold amplification, maintenance, and the global use of representations. It has not, however, been demonstrated that broadcasting alone is sufficient to explain the phenomenal quality of experience. In the remainder of this article, the comparison with MEM should therefore focus principally on whether MEM's proposed mechanisms of sensory content, re-entrant reconstruction, interoception, valence, and motivation provide additional, empirically discriminable explanations. At present, one may speak of a hypothesis of such an explanatory advantage, requiring quantitative tests, rather than a demonstrated replacement of GNWT by MEM.

## 3. The Motivated Emotional Mind Model: Origins, Architecture, and Mechanisms

The Motivated Emotional Mind (MEM) model is a developing physicalist programme that seeks to describe a common architecture for perception, memory, learning, emotion, thought, and action, while also identifying a mechanism of phenomenal consciousness. Its central proposal is to treat the brain, body, receptors, and effectors as a single regulatory system. External and internal information is recognized within a multilayer associative memory, evaluated in relation to the state of the organism, used to select an action, and then recurrently reconstructed in sensory and interoceptive fields. The same cycle and processes are proposed to underlie both conscious access and the qualitative, first-person character of experience (Galus, 2025a, 2025b, 2025c, 2026; Galus & Starzyk, 2020, 2026. Starzyk & Galus 2026).

Analysis of the sources requires a distinction among three levels of claims. The first comprises well-documented components incorporated into MEM: hierarchical sensory processing, recurrent activity, synaptic plasticity, distributed and sparse coding, interoception, and the influence of affect on learning and action. The second consists of MEM's integrative hypotheses, such as the organization of memory traces—engrams—into semblions and the identification of secondary perception with phenomenal availability of content. The third comprises more detailed implementation hypotheses, including functional couplings between postsynaptic membranes, a distinctive role for ephaptic coupling, astrocytes, or molecular memory coding. Components at the first level have independent support, whereas the complete architecture and its sufficiency for consciousness have not yet been directly confirmed experimentally.

### 3.1. Origins and Development of the MEM Model

The oldest foundation of MEM is the associationist tradition: learning binds co-occurring stimuli, responses, and states, so that activation of one element can reactivate another. In the twentieth century, this idea acquired a connectionist form. Hebb's rule provided a general principle for strengthening co-active connections (Hebb, 1949); Minsky's K-lines characterized memory as a structure capable of reinstating a previous state of mind (Minsky, 1980); Grossberg described resonance between ascending information and a descending memory signal (Grossberg, 1980); Felleman and Van Essen (1991) described a distributed hierarchy of visual areas; and Perlovsky (2001) proposed neural modelling fields and the dynamic matching of vague models to sensory data. Vadakkan (2010, 2011, 2012) proposed functional links between postsynaptic membranes as a possible mechanism generating a 'semblance,' that is, an internal semblance of previous activation. Architecture of Consciousness integrated these concepts into a model of a multilayer, semi-hierarchical network capable of categorization, generalization, and intra- and intermodal association (Galus, 2015a, 2015b).

A second line of development arose from theories of recurrent processing. Classical ascending flow allows features to be extracted and objects to be recognized rapidly, whereas descending connections permit completion, stabilization, recall, and imagery. The Recurrent Processing Theory of Lamme and Roelfsema (2000) distinguished an early feedforward sweep from subsequent recurrent processing. MEM adopted this distinction but assigned more specific functions to both directions: FFS is proposed to create and select representations and to guide learned action, whereas RP reconstructs states of lower sensory and interoceptive fields, thereby enabling secondary perception (Galus, 2025a, 2025b, 2025c, 2026).

A third line concerns embodiment, homeostasis, and affect. In the work of Damasio (2010), Panksepp (1998), and Solms (2021), the state of the organism is not an addition to cognition but a fundamental source of value and motivation. MEM combines this idea with the concept of motivated learning developed by Starzyk and colleagues (Starzyk et al., 2012): needs and violations of homeostatic or allostatic ranges generate affective pressure that alters attention, learning rate, representational selection, and action policy. Hierarchical memory is therefore not a neutral classifier. Its representations retain valence derived from the organism's individual history (Galus, 2025b, 2026; Galus & Starzyk, 2020, 2026).

## 3.2. Architecture and Morphology of the MEM System

The fundamental unit of analysis is not an isolated brain but an embodied agent. Exteroceptors provide signals about the environment; interoceptors and proprioceptors signal the state of the body; neuronal systems form memories and select responses; effectors alter the body or the environment; and the consequences of action become stimuli once again. This creates a closed causal loop: perception → recognition → valuation → action → renewed perception. In humans and animals, this loop is proposed to underlie both adaptive behaviour and individual perspective, because content is evaluated in relation to the needs of a particular organism (Galus, 2025a, 2025b, 2026; Galus & Starzyk, 2020, 2026a).

MEM morphology comprises successive levels extending from receptors and sensory fields, through feature-extracting fields and representations of objects and scenes, to high-level conceptual, premotor, and motor structures. The hierarchy is only partial. Lateral connections bind representations at the same level; cross-modal connections link vision, hearing, touch, smell, taste, proprioception, and interoception; and feedback projections run from higher to lower fields. A semi-hierarchical or heterarchical structure is therefore a more accurate description. It simultaneously permits upward compression of information, horizontal propagation of associations, and downward reconstruction of detail (Galus, 2015a, 2015b, 2018, 2022, 2025b; Galus & Starzyk, 2020).

A semblion is a postulated multilayer, dynamic associative structure. Its lower layers are anchored in maps of receptor activation; intermediate layers represent features, configurations, and subcategories; higher layers may correspond to objects, concepts, scenes, ideas, models of reality, relations, action programmes, and emotional contexts. A semblion is not a single 'concept cell.' A neuron may participate in multiple structures, while a small apical population may index an extensive, distributed trace at lower levels. This connects sparse with distributed coding and avoids identifying an entire representation with a single neuron (Galus, 2025b, 2022; Galus & Starzyk, 2020).

Semblions constitute memory and simultaneously serve as instruments of processing. Repeated co-activation strengthens transmission pathways; similarity between a new pattern and a consolidated trace facilitates reactivation; lateral connections trigger associations; and competition and inhibition select a dominant pattern. Dynamic semblions are additionally proposed to encode sequences: successive components of an episode or action programme are consolidated in temporal relations and may later be reinstated. The same architecture is therefore intended to support semantic, episodic, and procedural memory (Galus, 2015b, 2018, 2025b, 2022, 2026; Galus & Starzyk, 2020, 2026).

The formation of a semblion can be described as the gradual consolidation of relations among simultaneously active components of experience. Repeated activation of receptors and their corresponding sensory fields co-activates representations of features, objects, context, bodily state, and responses; plasticity stabilizes pathways through which later activation of part of the pattern can reinstate a larger whole. In the account by Galus and Starzyk (2020), an upper, relatively sparse population functions as an index, whereas content remains distributed across overlapping subsemblions in lower layers. The concept draws on Vadakkan's hypothesis that functional LINKs between activated postsynaptic membranes may evoke a 'semblance' of prior activation (Vadakkan, 2010, 2011, 2012). This is a candidate implementation, not a definitional condition of MEM or a mechanism already demonstrated for a complete percept.

The 2015 studies and later works identify several possible implementation mechanisms: Hebbian plasticity and LTP, neurogenesis and synaptogenesis, dendritic-spine dynamics, local protein synthesis and modification, epigenetic processes, tripartite synapses and astrocytic calcium waves, ephaptic coupling, and functional links between adjacent postsynaptic membranes (Galus, 2015a, 2015b; Galus & Starzyk, 2020). Aur and Jog's NeuroElectroDynamics (2010) is invoked here as the hypothesis that information may also be carried by the specific spatiotemporal organization of ionic currents, rather than solely by firing rate. These mechanisms should not be presented as the jointly demonstrated substrate of the semblion. They are candidate implementations with unequal levels of support; MEM can retain its architectural level even if some detailed molecular hypotheses prove incorrect.

MEM assigns a distinctive role to a parallel interoceptive pathway. Disruptions of organismic equilibrium are registered as hunger, pain, tension, thirst, dyspnoea, or other bodily signals. These states modulate representational selection, plasticity, and action selection. Descriptions of the model include a 'decision triangle' comprising the superior colliculus, periaqueductal grey, and mesencephalic locomotor region: sensory, affective, and motor systems are proposed to acquire a shared action context within this circuitry (Galus, 2025b, 2025a, 2026; Galus & Starzyk, 2020). This is a useful systems-level hypothesis, but it should not be interpreted as an established, self-contained locus of decision-making or consciousness.

### 3.3. A Common Mechanistic Scheme

The unity of MEM lies in the repeated application of the same cycle to different contents and timescales. It does not reduce all phenomena to a single synapse or one type of activity. Rather, it is a common organizational scheme in which the following stages occur in sequence:

1. State registration. Exteroceptors, interoceptors, and proprioceptors generate modality-specific activity patterns concerning the environment and the organism.
2. Ascending FFS processing. Activity passes through successive levels; features are extracted, and new configurations are compared with traces of previous learning.
3. Categorization and selection. Similarity to consolidated patterns, priming, attention, inhibition, and WTA competition give rise to dominant semblions.
4. Association and valuation. Lateral and cross-modal connections add mnemonic, interoceptive, emotional, and linguistic context, together with possible action programmes.
5. Working access and action. Active apical layers influence working memory, speech, planning, and motor fields; in rapid responses, this stage may proceed without phenomenal awareness.
6. Recurrent process (RP). Activity from selected representations returns to lower sensory and interoceptive fields, reconstructing parts of the pattern associated with previous perception.
7. Secondary perception. The reconstructed pattern is processed again as a perceptual signal; according to MEM, this provides the basis for mental imagery, feeling, and phenomenal awareness of content.

8. Updating and the next cycle. The outcome modifies memory, valence, expectations, and readiness for action; a renewed FFS may develop an association into a thought, scenario, utterance, or new decision.

Phenomenal experience requires not only the global use of a representation, but also anchoring in modality-specific fields and the participation of an interoceptive–affective state. Interoception need not imbue every content with strong emotion, but it specifies the organism's perspective and the content's regulatory significance. This claim is stronger than the thesis of recurrence alone and constitutes one of MEM's principal empirically risky assumptions (Galus, 2025a, 2025c, 2026; Galus & Starzyk, 2026a).

## 3.4. Principal Mental Processes Explained by MEM

### 3.4.1. Formation of Representations

A representation arises through consolidation of recurrent or particularly significant configurations of sensory activation. Its content is not stored in a single location: it comprises a modality-specific trace, higher-order features, relations to other modalities, bodily context, valence, and possible responses. Learning thus forms an engram with a semi-hierarchical structure. Sparse coding at the level of concept neurons serves as an index or apical neuron, whereas the complete content remains distributed across subsemblions and lower fields (Galus, 2015b, 2022; Galus & Starzyk, 2020).

During the ascending FFS, perceived objects are categorized and generalized. Categorization involves matching the current configuration to consolidated patterns and selecting the best match. Generalization proceeds towards representations of shared features, whereas discrimination and identification of an individual object draw on more detailed substructures. Horizontal associations link co-occurring objects, and cross-modal connections bind image, sound, touch, action, and emotion. The result is a model of the scene and a broader model of reality, rather than merely a classificatory label (Galus, 2015a, 2015b, 2025b, 2022; Galus & Starzyk, 2020).

### 3.4.2. Perception

MEM distinguishes direct receptor-based perception, ascending recognition, and secondary reconstruction. Lower maps preserve the modality-specific structure of the stimulus; FFS extracts features and links them to memory; RP completes or reinstates the pattern. A percept results from the interaction of current data with learned representations and is not a passive copy of the stimulus. In this account, phenomenality is not added only at the stage of abstract report, but depends on activation or reactivation of fields anchored in receptors, together with a concurrent interoceptive context (Galus, 2025b, 2025c, 2026).

### 3.4.3. Thought and Access and Phenomenal Consciousness

Thought is a sequence of semblion activations initiated by a stimulus, an association, a knowledge gap, mental saccades, or a state of need. Content can move laterally between representations or vertically among their levels. Activation of apical fields makes a concept available to working memory, speech, reasoning, and action, thereby implementing the functions of access consciousness. When this activity triggers recurrent sensory and interoceptive reconstruction, the content is proposed to acquire the form of an experienced image, sound, tension, or emotion. MEM thus relates access and phenomenality to successive phases of the same loop without mechanically identifying them (Galus, 2025b, 2025a, 2026).

### 3.4.4. Memory, Imagery, Dreaming, Hallucination, and Illusion

In MEM, the differences among these phenomena arise primarily from the source and relative contribution of activation. In external perception, receptor data predominate and are corrected by re-entry. In memory or imagery, top-down activation originating in a consolidated semblion predominates. Sleep reduces ongoing sensory constraint and permits freer reconstruction. Hallucination would occur when an internal pattern acquires a status resembling that of an external signal despite the absence of an adequate stimulus; illusion occurs when an actual signal is incorrectly completed or categorized by memory. In every case, the basic mechanism remains the same: top-down reactivation and renewed perception of modality-specific fields (Galus, 2025a, 2025b, 2024, 2026).

### 3.4.5. Language and Inner Speech

The language model is based on associations between representations of objects, actions, and relations and concurrently heard or produced phonemes and graphic signs. A word semblion therefore links an auditory or visual layer with a multimodal concept and action schemas. Structural similarities among representations facilitate metaphor and analogy, while dynamic sequence semblions may encode phrases and rules of use. The phonological loop, linking auditory and articulatory areas, maintains inner speech. Semantics is grounded in experience, whereas syntax emerges from consolidated sequences and relations among symbols (Galus, 2018, 2026).

### 3.4.6. Brain–Mind Identity and Causation

In MEM, a mental state is not a separate entity correlated with a brain state, but a mode of operation of a specific brain–body–receptor configuration. The causation of action is implemented by the ascending and motor physical pathways: stimulus, memory, and needs select a programme that activates effectors. Phenomenal report is a subsequent reconstruction of what the system perceives, remembers, or plans. It may be locally epiphenomenal with respect to an event that has already occurred, but is not globally causally inert, because reconstructed content modifies subsequent cycles of attention, memory, and action. Distinguishing the executive from the reporting phase is intended to reconcile the causal closure of physics with the experience of agency (Galus, 2024, 2025a, 2026).

### 3.4.7. Understanding, Curiosity, and the Mathematical Model

Understanding is conceived as incorporating recognized content into a hierarchy of knowledge and linking it to context, consequences, and possible actions. Symbolic similarity is insufficient: a representation must be grounded in sensory, interoceptive, and motor history. The mathematical formalization (Galus & Starzyk, 2026a) describes the system state in terms of cognitive–sensory representations, bodily–motivational context, allostatic variables, and affect. Semblions compete on the basis of similarity, context, and associative activation; an affective signal modulates learning rate and valence; and a policy selects programmes that minimize the predicted cost of regulatory violations. A discrepancy between the ascending signal and reconstruction produces a novelty error which, together with affective pressure, increases curiosity and exploration. When no known programme attains sufficient value, a procedural gap arises and initiates the generation and simulation of new programmes—a candidate mechanism of creativity (Galus & Starzyk, 2026a, Starzyk & Galus 2026).

### 3.4.8. Phenomenal Consciousness

In MEM, phenomenal consciousness is not an additional effect arising above neuronal and biophysical processes, but is identified with a specific regulatory–perceptual state of the system. This state comprises secondary activation of lower exteroceptive and interoceptive fields and its renewed perceptual processing. In the cycle: direct perception → FFS → RP → secondary perception, modality-specific qualia are linked to concurrently active interoceptive representations, which in MEM jointly constitute the affective component of experience. This account draws on the James–Lange theory (James, 1884; Lange, 1885/1922), according to which complexes linking interoceptive and cognitive states are identical with innate or learned affective states. Phenomenal consciousness is thus, in this model, a holistic perceptual state comprising modality-specific content, its bodily significance, and the current regulatory–affective context.

### 3.4.9. Continuity between Human and Animal Brains

MEM assumes continuity of mechanism across species. Non-linguistic animals can form multimodal representations, learn valences, select actions, reinstate episodes, and simulate simple scenarios. Differences concern the number of levels, richness of associations, planning horizon, episodic memory, and capacity for symbolization, rather than the emergence of an entirely new principle in humans. The simplest receptor–motor responses are nevertheless insufficient for attributing full consciousness in the MEM sense; at minimum, plastic associative memory, need-dependent selection, and recurrent reinstatement of content are required. The boundary remains a graded hypothesis and requires comparative operationalization.

### 3.5. Scope of Unification and Principal Limitations of MEM

MEM explicitly encompasses a broader range of phenomena than standard formulations of GNWT: direct perception, recognition, automatic action, conscious access, mental imagery, memory, dreaming, hallucination, emotion, inner speech, planning, curiosity, and creative generation of programmes. Their common denominator is an embodied, affectively regulated associative–recurrent loop. In this precise sense, MEM can be said to propose a single coherent organizational mechanism for access-related and phenomenal processes. This does not imply, however, that all these phenomena have identical implementations, time courses, locations, or causal dependencies.

The programme's advantage lies in relating content to its modality-specific vehicle, valence, and organismic needs, and in providing a unified account of perception, memory, action, and endogenous content. This permits the formulation of mathematical variables and empirical tests without restricting consciousness to report.

The limitations are equally important. A semblion has not been demonstrated as a distinct neurobiological unit spanning the complete pathway from receptors to action. Evidence concerning engrams, concept cells, sensory reactivation, and plasticity supports components of the hypothesis but not its complete geometry. It has likewise not been established whether reactivation of early fields and parallel interoception are necessary for every type of conscious content. NED, ephaptic, astrocytic, and molecular hypotheses must be carefully distinguished from well-supported synaptic plasticity.

The proposed mathematical formalism (Galus & Starzyk, 2026a) is an important step, but remains a general framework rather than a complete generative model. It does not yet unambiguously specify all functions, parameters, and timescales; provide stability analysis or fit to data; resolve identifiability; or bridge mathematical activity and phenomenal quality without relying on an interpretive hypothesis. Particularly needed are experiments that dissociate the predictions of MEM and GNWT, preregistered falsification criteria, and simulations showing that a single parameter set can reproduce behaviour, report, affective changes, and neuroimaging properties.

## 4. The Mechanistic Relationship between MEM and GNWT

### 4.1. From the Global Workspace to an Embodied Network of Semblions

Replacing GNWT mechanisms should be understood as substituting explanatory functions, not as a demonstrated refutation of the theory. GNWT explains conscious access through competition among local representations, threshold ignition, recurrent maintenance, and the long-range availability of the winning content to multiple cognitive systems (Dehaene, Kerszberg, & Changeux, 1998; Dehaene & Naccache, 2001; Mashour et al., 2020). MEM proposes a different unit of organization: not a separate global workspace, but an embodied, semi-hierarchical network of semblions extending from receptor fields through feature and cognitive fields to working memory, procedural structures, and effectors. A semblion is both a representation and a processing pathway; it can link a modality-specific trace, its compressed category, associations, valence, and a possible action programme (Galus & Starzyk, 2020; Galus, 2022, 2024).

During FFS, multiple configurations of stimulation compete for further transmission. Match to consolidated patterns, prior priming, the strength and extent of activation, and lateral inhibition give rise to winner-take-all (WTA) dynamics. Successive layers retain recurrent and significant features while discarding much of the detail. Experience is thereby compressed: from elementary signal differences and object features, through categories and generalizations, to ideas and a model of the environment. The winning pattern is therefore not an arbitrarily selected message, but activation of a structure shaped by the learning history of the individual organism (Galus, 2015a, 2015b; Galus & Starzyk, 2020). Functionally, WTA replaces competition for access in GNWT, while hierarchical compression explains why generalized, symbolic content reaches working memory while remaining grounded in receptor-level details and qualia.

### 4.2. Perception, Qualia, and Motivation as Components of the Same Architecture

In MEM, direct perception is based on differentiating receptor signals and incorporating them into learned object patterns. FFS extracts features, matches them to memory, categorizes, and generalizes, thereby forming a cognitive model of the scene. Because a semblion includes both lower receptor fields and higher cognitive fields, a recognized concept retains access to the modality-specific trace from which it arose. On this account, a quale is not an additional label appended within the global workspace, but the qualitative aspect of activity in receptors and modality-specific sensory fields, consolidated through the body's individual interactions with the environment. The book by Galus and Starzyk (2020) and later articles develop this thesis, arguing that properties such as slipperiness, pressure, colour, or taste acquire meaning by associating sensation with manipulation, the outcome of action, and learning history (Galus, 2024, 2025a, 2025c).

Semblion associations can link percepts from exteroceptors with concurrent percepts from interoceptors and proprioceptors. Deviations from homeostasis—hunger, thirst, pain, dyspnoea, tension, or other regulatory disturbances—are registered by interoceptive systems and jointly constitute affective states. In MEM, these states are not passive commentary on cognition: they alter priming strength, the course of WTA competition, learning, and response selection. An emotional state therefore becomes a motivation for action, while successful responses are consolidated in the motor branches of semblions as procedural memory. Recurrence of a similar external and internal configuration may automatically trigger a learned behavioural programme (Galus & Starzyk, 2020; Galus, 2024, 2026; Galus & Starzyk, 2026a). This mechanism replaces neutral broadcasting with a richer process: the representation made available is embodied, valued, and linked to a repertoire of actions from the outset.

### 4.3. Secondary Perception, Endogenous Contents, and Report

Re-entry proceeds from active cognitive fields towards lower sensory and interoceptive fields. The reinstated modality-specific pattern is then recognized again during a subsequent FFS; MEM refers to this stage as secondary perception. When re-entry originates in memory or associative activity, it produces visualization of a memory, imagined scene, or planned action. With diminished control by receptor input, a similar mechanism is proposed to generate dreams; when endogenous activation wrongly predominates, it produces hallucinations and some illusions. Recurrent reconstruction may also encompass interoceptive fields, so a recalled image or scenario may recover part of its former emotional charge (Galus, 2024, 2025a, 2025c, 2026). Neuroimaging evidence of reactivation in early sensory fields during imagery is compatible with this hypothesis, although it does not yet confirm the complete MEM architecture (Slotnick, Thompson, & Kosslyn, 2005).

Standard GNWT explains when the content of a memory or image becomes globally available, but does not offer a specific mechanism that generates the modality-specific form of imagery, memories, and dreams. Such phenomena can be described as contents activated and broadcast within the workspace, but broadcasting itself does not determine the source of their visual, auditory, bodily, or emotional character. MEM addresses this gap through a mechanism of top-down decompression: a compressed cognitive representation reactivates fragments of its own modality-specific fields, and their renewed perception, aggregated with associated interoceptive states, forms phenomenal content. This does not mean that GNWT is logically incapable of incorporating generative mechanisms, but rather that such mechanisms do not belong to its classical explanatory core (Dehaene & Naccache, 2001; Mashour et al., 2020).

Report arises when the reconstructed content is recategorized and linked to linguistic representations and programmes for speech or other communicative action. Cross-modal associations bind appearance, sound, touch, and action to phonemes, words, gestures, or signs, thereby enabling the learning of speech and other communication systems. Inner speech can maintain and organize sequences of thought, whereas overt speech activates procedural articulatory programmes (Galus, 2018, 2026; Galus & Starzyk, 2020). MEM thus replaces GNWT's report function with a specified sequence: re-entry → secondary perception → association with a linguistic semblion → activation of a motor programme.

### 4.4. Scope of Complete and Incomplete Substitution

MEM proposes counterparts for the five GNWT functions considered here: selection, working memory, maintenance, report, and action control. WTA selection and the transition to learned action are described most fully; global ignition, scope of availability, working-memory gating, and flexible routing require further operationalization. The model additionally introduces hierarchical compression, connections between receptor and cognitive fields, interoceptive–affective valuation, procedural learning of responses, and secondary perception of endogenous contents. Table 1 distinguishes existing MEM mechanisms from proposed extensions.

The greatest difference concerns phenomenality. GNWT is primarily a theory of global availability, whereas MEM locates qualitative content in modality-specific sensory–interoceptive reconstruction. If disruption of such reconstruction abolishes or specifically alters experience while broadcasting and task performance remain intact, MEM will have gained evidence for a mechanism extending beyond GNWT. If, however, global access proves sufficient and conscious content is preserved without the postulated modality-specific reactivation, the central MEM claim will be weakened.

The comparison must nevertheless address contemporary, multilevel GNWT. Its recent models relate global dynamics to the connectome, AMPA/NMDA receptors, development, and valence (Klatzmann et al., 2025; Changeux & Farisco, 2026). MEM therefore gains no advantage merely by invoking biology, the body, or emotion. The hypothesis of an explanatory advantage can be supported only by demonstrating that the characteristic coupling of semblion compression, interoception, procedural memory, and secondary perception predicts findings that multilevel GNWT cannot reproduce without adding mechanisms outside its own core.

**Table 1. Functional substitution of GNWT mechanisms by MEM mechanisms. The assessment concerns correspondence in explanatory role, not equal strength of empirical support.**

| GNWT function | MEM mechanism | Assessment of substitution | Gap and proposed addition |
|---|---|---|---|
| **Selection** | Competition among FFS configurations, dependent on similarity, priming, activation strength, lateral inhibition, and WTA dynamics; interoceptive–affective modulation. | High at the functional level; unverified as mechanistic equivalence | Empirically established thresholds and temporal dynamics are lacking. A mathematical model could define a matching function, gain, and a context-dependent winning threshold (Galus & Starzyk, 2026a). |
| **Working memory** | A temporarily dominant configuration of the apical fields of a semblion, available to associations, language, planning, and action. | High at the functional level; partial at the mechanistic level | The neurobiological input/output gate and the capacity of the mechanism have not been specified. A candidate addition is reversible amplification of the winning semblion by corticothalamic loops, priming, and an affective signal. |
| **Content maintenance** | Repeated FFS–RP cycles, priming, and reactivation of distributed semblion components. | Partial | It remains unclear when persistent activity is necessary and when a latent activity trace is sufficient. MEM could combine active re-entry with short-term changes in excitability or connection strength and predict distinct markers for the two modes. |
| **Global ignition** | Threshold victory of a widespread representation and transition to a self-reinforcing FFS–RP loop. | Partial; no evidence of equivalence | This mechanism has not been shown to be equivalent to GNWT ignition. A possible addition would be a formally defined bifurcation: after the threshold is crossed, the dominant semblion rapidly increases its spatial extent and duration of activation (cf. Klatzmann et al., 2025). |
| **Global availability** | Propagation of activation through overlapping semblions, lateral and cross-modal connections, and linguistic, mnemonic, and motor branches. | Partial; scope has not been operationalized | Rich associativity does not guarantee global availability. MEM should define the functional set of recipients, a measurable index of the winning semblion's reach, and the capacity to route content rapidly to novel task systems. |

| GNWT function | MEM mechanism | Assessment of substitution | Gap and proposed addition |
|---|---|---|---|
| **Report** | Secondary perception, recategorization, coupling to linguistic semblions, and activation of a procedural communication programme. | High at the functional level | An explicit threshold of reportability and a model of metacognitive confidence are lacking. They might be derived from the strength and stability of reconstruction, agreement across successive cycles, and the degree of activation of the linguistic representation. |
| **Action control** | WTA selects a representation linked to a procedural programme; re-entry reconstructs possible outcomes, while interoception and valence differentiate scenarios. | High for selection; partial for flexible control | Conflict, response inhibition, and rule switching remain underspecified. A candidate addition is multicycle competition among simulated scenarios according to predicted allostatic cost and the procedural gap (Galus, 2024, 2026; Galus & Starzyk, 2026a). |
| **Phenomenal content** | Receptor-grounded activation and recurrent reconstruction of modality-specific sensory and interoceptive fields, leading to secondary perception. | **Not substitution, but an extension beyond the classical GNWT core** | This is a distinctive but as yet unverified MEM claim. It requires experimentally separating the effect of re-entry on experiential quality from its effects on attention, working memory, report, and task performance. |

*Note. FFS—feedforward sweep; RP—recurrent processes/re-entry; WTA—winner-take-all mechanism. The proposed additions are hypotheses for the further development of MEM, not components that have already been confirmed in direct comparative tests.*

### 4.5. Shared, Orthogonal, and Genuinely Competing Claims

GNWT and MEM are not straightforward opponents operating at a single level of description. Some claims are shared: both accounts assume distributed processing, competition among representations, a nonlinear transition to a more stable state, recurrence, and the capacity of the winning content to influence multiple downstream processes. Other claims are orthogonal. GNWT specifies the conditions and dynamics of global availability, whereas MEM describes the geometry of learned representations, their receptor grounding, their relations to action programmes, and their affective valuation. These components may coexist in the same brain.

Genuine competition begins only where the theories assign a constitutive role to different processes. Strong GNWT holds that sufficiently rich ignition and global availability are sufficient for conscious experience. Strong MEM holds that modality-specific sensory–interoceptive activation or reconstruction is necessary for phenomenal quality. An experiment can therefore adjudicate between the theories only if it independently manipulates global availability and modality-specific re-entry while measuring experience in a manner as free as possible from reporting demands.

GWT should also be distinguished from GNWT. The functional metaphor of a global workspace does not determine its implementation, whereas GNWT includes specific hypotheses concerning long-range projection neurons, cortical layers, AMPA and NMDA receptors, excitation–inhibition balance, recurrent maintenance, and dynamical bifurcation. Replacing the term 'broadcast' with the general term 'propagation' is therefore insufficient to replace GNWT. MEM must predict which populations, transmission directions, delays, and thresholds will be observed and when they will differ from the predictions of a global workspace.

## 5. Method for Comparing the Empirical Evidence

A major factor placing GNWT among the most prominent theories of consciousness is the extensive body of experiments and models addressing its predictions. To determine the position of MEM relative to GNWT, we examined the extent to which the MEM paradigm is compatible with the findings of these studies. A successful *ex ante* prediction, formulated before the experimental results are known, has greater evidential and confirmatory value than a post hoc explanation constructed after those results have become available. Nevertheless, such a post hoc reinterpretation may justify further comparative investigation of theories based on different ontological and structural assumptions. We therefore selected a corpus of 12 publications concerning GNWT and its critique: empirical studies, computational models, and one influential theoretical paper. Because this corpus does not address interoception, to which MEM assigns an important role, we added 13 publications on bodily signals, affect, and organismic regulation. These are influential studies by reputable research groups, but their findings—like all empirical evidence—remain open to methodological assessment and competing interpretations. The following subsections present the comparison criteria, results, and resulting conclusions.*ex ante*
The corpus comprises 25 publications. Items 1–11 present experiments and models commonly cited in support of GNWT; item 12 represents an influential recurrent-processing critique. Items 13–25 concern interoception, affect, heart–brain coupling, respiration, facial expression, and cortical stimulation. The second group was not designed as a test of MEM and cannot be treated as direct confirmation of the theory. It was included because it examines variables to which MEM assigns a central role and which classical GNWT addressed less extensively (Barrett & Simmons, 2015; Seth & Friston, 2016; Parr, Pezzulo, & Friston, 2022; Pezzulo, Parr, & Friston, 2024).

The research groups and journals considered are highly reputable, but the designation ‘canonical’ refers to the influence of these studies on theory formation, not to immunity from criticism. Measures of consciousness, reporting requirements, the relation between correlation and causation, the interpretation of P3b and frontal activity, and the question whether markers of access are also markers of phenomenality remain disputed. We therefore compare the data at the operational level rather than relying on the authors' stated conclusions.

Sixteen criteria across four dimensions were used. Each publication was rated on a 1–6 scale: 6—strong support, 5—support, 4—weak support, 3—neutral, 2—weakening evidence, and 1—strong contradiction. Ratings concern the data or model properties in relation to the criterion, not the prestige of the publication. Empirical studies and meta-analyses received a weight of 1; hybrid models with biological validation, 0.75; computational models, 0.5; and theoretical papers and reviews, 0.25. These weights are explicit but conventional; future sensitivity analyses should test the stability of the results under alternative weightings.

**Table 2. Theory-neutral comparison criteria.**

| Code | Criterion |
|---|---|
| **P1** | Transition from neuronal activity to experienced content |
| **P2** | Modality-specific, qualitative, and affective character of experience |
| **P3** | Unity, stability, and temporal extension of experience |
| **P4** | Experience independent of immediate report; endogenous contents |
| **A1** | Selection, competition, and limited capacity |
| **A2** | Threshold behaviour and transition dynamics |
| **A3** | Integration, access, and maintenance |
| **A4** | Flexible use of content in report, memory, and action |
| **S1** | Speed and economy of processing |
| **S2** | Context, uncertainty, and robustness |
| **S3** | Learning and adaptation |
| **S4** | Goals, action, and organismic regulation |
| **B1** | Hierarchy and feedback |
| **B2** | Cortical and subcortical networks |
| **B3** | Temporal and electrophysiological correspondence |
| **B4** | Causation, no-report paradigms, and cross-species comparisons |

Three quantities were determined for each criterion. M denotes the theory's stated explanatory scope: how broadly and specifically the theory covers a given criterion. E denotes the weighted consistency of the examined corpus with that criterion, independently of either theory. K = min(M, E) is a conservative coefficient bounded by the weaker of the two components. M is not a measure of truth or of fit to a single experiment; a high M may reflect an ambitious theoretical scope not yet fully supported by evidence. Nor is E a measure of causal strength. K is intended to prevent a broad theoretical claim from compensating for insufficient evidence, or a rich body of evidence from attributing to a theory a scope it does not claim.

The numerical results constitute a structured synthesis rather than a meta-analysis. Ratings depend on publication selection and interpretation of the criteria. The interoceptive group in particular increases the visibility of the strengths of MEM and interoceptive predictive processing, and is therefore also presented separately. Our conclusions rest principally on the pattern of correspondence and on discriminating possibilities, rather than on differences of hundredths of a point.

## 6. Canonical GNWT Studies and Their Reinterpretation in MEM

The analyses below are presented in no particular order. In each case, we distinguish the observed result, the standard GNWT interpretation, a possible MEM interpretation, and diagnostic value. Compatibility of MEM with a finding means that the architecture can reproduce it qualitatively; it does not imply that the result was predicted in advance using parameters independent of the study.

**1. The Attentional Blink and the Discreteness of Access**
**Source: Sergent and Dehaene (2004)**

Study and result. In an attentional-blink task, a second target presented shortly after the first often fails to reach awareness. The distribution of subjective visibility was more consistent with a discrete transition than with a gradual reduction in quality.

GNWT interpretation. The first target occupies the limited workspace; the second fails to reach the threshold for neuronal ignition while recurrent amplification resources remain engaged.

MEM interpretation and discriminating value. The first winning semblion is primed and maintained by WTA and RP. The second signal may undergo an early FFS but fail to establish a stable recurrent loop and secondary perception. The result supports threshold selection shared by both theories. Adjudication would require predicting the complete temporal curve, including lag-1 sparing, from independently estimated parameters.

**2. Temporal Markers of Conscious Access during the Attentional Blink**
**Source: Sergent, Baillet, and Dehaene (2005)**

Study and result. MEG showed that seen and unseen stimuli differ in late, widespread activity, whereas early sensory responses may remain similar.

GNWT interpretation. The late divergence corresponds to ignition of the global network and broad availability of the content following preserved early processing.

MEM interpretation and discriminating value. MEM distinguishes local FFS from the later victory of a representation, recurrent reconstruction, and associative dissemination. The late timing of the effect is therefore also predicted by the FFS–RP cycle. Chronometry alone identifies neither the recipients of broadcasting nor the modality-specific target of re-entry. Directionality measures and layer-specific interventions are required.

**3. Intracranial Markers of Conscious Access**
**Source: Gaillard et al. (2009)**

Study and result. Masked words elicited brief, relatively local responses; consciously seen words were associated with sustained potentials, high-frequency activity, phase synchronization, and long-range connectivity.

GNWT interpretation. The convergence of multiple late markers is the classical signature of neuronal ignition and stable entry of content into the global workspace.

MEM interpretation and discriminating value. In MEM, a weak signal may activate only fragments of the FFS pathway, whereas a winning semblion triggers RP and propagates through linguistic, semantic, and motor associations. This is strong evidence for widespread recurrent dynamics in conscious access, but not for the exclusivity of the GNWT architecture. MEM must quantitatively derive the timing, frequency bands, and topology of the effect.

**4. Local and Global Auditory Regularities**
**Source: Bekinschtein et al. (2009)**

Study and result. In the local–global paradigm, the brain responded to local violations of regularity even without full consciousness, whereas violations of a higher-order rule were associated with a late, widespread response dependent on the state of consciousness.

GNWT interpretation. Representing the global rule requires integration of a sequence within the workspace and its availability to multiple systems.

MEM interpretation and discriminating value. A dynamic sequence semblion can encode temporal relations; violation of a higher-order pattern requires multicycle integration, re-entry, and secondary perception of the mismatch. The study distinguishes levels of automatic and conscious processing, but both models predict such a distinction. Selective disruption of global routing with preserved modality-specific reconstruction, or the reverse dissociation, would be diagnostically informative.

**5. Prefrontal Cortex Lesions and the Threshold of Subjective Visibility**
**Source: Del Cul et al. (2009)**

Study and result. Patients with prefrontal cortex lesions exhibited an elevated threshold for subjective visual access despite relatively preserved objective processing.

GNWT interpretation. The result has causal significance: the prefrontal component of the global network participates in threshold crossing and conscious access.

MEM interpretation and discriminating value. MEM can treat prefrontal cortex as a high-level component of the active semblion that is important for WTA, metacognition, decision-making, and report. A lesion may impair selection or readout without abolishing all modality-specific components of the percept. The study poses a greater challenge to MEM if experience disappears despite preserved sensory RP and no-report measures. Without this dissociation, it does not determine whether the quale or the reporting criterion changed.

**6. Internally Generated Perceptual Transitions without a Change in the Stimulus**
**Source: Kapoor et al. (2022)**

Study and result. During binocular rivalry, prefrontal activity nonlinearly discriminated spontaneous changes in perceptual content, including under procedures designed to reduce dependence on report.

GNWT interpretation. This supports a role for prefrontal cortex in conscious access rather than exclusively in preparing a motor response.

MEM interpretation and discriminating value. MEM interprets prefrontal activity as a shift in dominance of a high-level semblion coupled to reconstruction of the corresponding sensory maps. The transition can arise endogenously because competition does not require a change in the stimulus. The data are compatible with both theories. Temporal order would be decisive: does a change in modality-specific RP precede the frontal signal, or does global availability initiate reconstruction?

**7. The Report Threshold: Signal Loss and Response Bias**
**Source: van Vugt et al. (2018)**

Study and result. Recordings from visual and frontal cortex in macaques dissociated loss of sensory information from decision bias: the visual signal constrained detection, whereas frontal activity was more strongly associated with the reporting criterion.

GNWT interpretation. GNWT can distinguish the quality of the input representation from frontal access, decision-making, and flexible use.

MEM interpretation and discriminating value. MEM describes an analogous division in terms of FFS strength/match in a modality-specific representation and high-level competition, valence, and response-programme selection. The finding does not simply confirm a single centre of consciousness; rather, it requires multistage models. MEM needs an explicit function relating semblion strength to the reporting criterion.

### 8. Cortical and Subcortical Structures in Conscious Perception without Immediate Report
**Source: Kronemer et al. (2022)**

Study and result. A multimodal study found widespread cortical and subcortical correlates of visibility, including dynamics independent of overt report or eye movements.

GNWT interpretation. The finding is compatible with the distributed, multilevel character of global access and argues against purely local interpretations.

MEM interpretation and discriminating value. MEM expects the involvement of sensory, associative, thalamic, and other subcortical structures because a semblion forms part of an embodied loop of selection, action, and valuation. The broad distribution supports both architectures. Adjudication requires determining whether subcortical signals carry affective context necessary for content, or merely modulate the probability of global ignition.

### 9. A Connectome- and Receptor-Constrained Model of Cortical Bifurcation
**Source: Klatzmann et al. (2025)**

Study and result. A whole-cortex model constrained by the connectome and AMPA/NMDA receptor gradients generated a bifurcation between rapidly decaying activity and widespread, sustained activity resembling ignition.

GNWT interpretation. The study strengthens the biological feasibility of one of GNWT's central mechanisms and links receptor-level properties to whole-cortex dynamics.

MEM interpretation and discriminating value. MEM can regard the same bifurcation as an implementation of the threshold victory of a semblion and the transition to a self-reinforcing FFS–RP loop. The model demonstrates mechanistic feasibility, not the identity of ignition with consciousness. To generate distinctive predictions, MEM would have to add representational geometry, the directionality of reconstruction, and interoceptive variables.

### 10. Spontaneous Activity and Inattentional Blindness
**Source: Dehaene and Changeux (2005)**

Study and result. A network model showed how fluctuations in spontaneous activity and competition can cause the same stimulus to gain access on some trials but not others.

GNWT interpretation. Small differences in the initial state determine whether the nonlinear threshold for global ignition is crossed.

MEM interpretation and discriminating value. In MEM, the initial state includes semblion priming, current context, inhibition, activation history, and allostatic pressure; these variables can likewise shift the outcome of WTA competition. This demonstrates feasibility and fit for a class of mechanisms, not a unique test. The models must be compared with the same parameter budget and on out-of-sample data.

### 11. A Network Model Linking Subjective Report to Physiological Data
**Source: Dehaene, Sergent, and Changeux (2003)**

Study and result. The model reproduced the difference between objective processing and subjective availability under masking and attentional constraints.

GNWT interpretation. A recurrent, long-range network provides a formal bridge from local processing to a global, reportable state.

MEM interpretation and discriminating value. Attractor competition and recurrence can also be implemented in a network of semblions. MEM adds an interpretation in terms of modality-specific re-entry and affective context. Formal agreement between the dynamics and the data does not establish a unique biological mechanism. Comparison requires a shared dataset, complexity penalties, and prospective predictions.

### 12. Recurrent Processing Theory as a Critique of Identifying Consciousness with Access
**Source: Lamme (2006)**

Study and result. The review argued that local recurrence in sensory cortex may constitute visual experience before global access and report.

GNWT interpretation. For GNWT, this is a competing interpretation that separates phenomenality from subsequent workspace functions.

MEM interpretation and discriminating value. MEM shares the emphasis on recurrent processes but adds the learned structure of the semblion, modality-specific reconstruction, interoception, and secondary perception. It is therefore not identical to RPT. The paper is not an empirical test of MEM, but it identifies the central experimental axis: sensory recurrence versus frontoparietal availability.

## 7. Studies of Perception, Interoception, and Affect

The interoceptive group extends the analysis beyond GNWT's classical domain. MEM predicts that bodily states and their neuronal representations jointly constitute the context, valence, and motivational significance of percepts. Contemporary active inference and interoceptive predictive processing advance related, although mathematically and conceptually distinct, claims: affect may arise from inference about the causes of bodily signals and from regulation of predicted allostatic states. Favourable findings in this group are therefore not the exclusive 'property' of MEM.

**13. Increased Heart Rate as a Cause of Anxiety-like Behaviour**

**Source: Hsueh et al. (2023)**

Study and result. Optogenetically induced tachycardia in mice increased anxiety-like behaviour in a specific context; the heart–insula axis was implicated.

GNWT interpretation. Multilevel GNWT can treat a visceral signal as one content or modulatory influence gaining global access, but the classical core does not predict the detailed heart–insula–behaviour relation.

MEM interpretation and discriminating value. MEM treats an interoceptive change as a component of an affective state that modifies WTA competition and action selection; context dependence corresponds to an association between a situational semblion and bodily valence. The result supports a causal role for interoception, but does not adjudicate between MEM and active inference or establish the necessity of this pathway for every quale.

**14. A Multicentre Test of the Facial-Feedback Hypothesis**

**Source: Coles et al. (2022)**

Study and result. In the Many Smiles Collaboration, some facial-expression manipulations produced small, procedure-dependent changes in self-reported happiness.

GNWT interpretation. GNWT can describe the change as the influence of a somatic signal on reportable content without providing a detailed model of valence.

MEM interpretation and discriminating value. MEM predicts that the proprioceptive–interoceptive trace of an action can alter the valuation of an active representation; the small effect is consistent with one among multiple contextual inputs. Methodological heterogeneity and the small effect size argue against the strong claim that facial configuration alone determines emotion.

**15. A Meta-analysis of Facial-Feedback Effects**
**Source: Coles, Larsen, and Lench (2019)**

Study and result. The meta-analysis found small and variable effects of facial expression on emotional experience, dependent on moderators and induction method.

GNWT interpretation. The data are compatible with the global use of bodily signals, but do not address the specific mechanisms of ignition.

MEM interpretation and discriminating value. For MEM, facial expression is a component of the body–brain loop, but its influence should depend on active context, learning history, and competing interoceptive signals. The result supports a moderate account, not strong somatic determinism.

**16. Cardiac-Cycle Phase and Detection of Threat Signals**
**Source: Garfinkel et al. (2014)**

Study and result. Fear-related stimuli were processed differently depending on cardiac-cycle phase, and the effect was associated with activity in interoceptive and emotion-related regions.

GNWT interpretation. GNWT can treat cardiac phase as modulating excitability or access for threat-related content.

MEM interpretation and discriminating value. MEM predicts coupling between the exteroceptive semblion of the stimulus and the momentary interoceptive state, which changes valence, priming, and the selection threshold. Evidence of an effect on experiential quality independently of the decision criterion would be crucial for MEM; the current data are compatible but nonspecific.

**17. Panic with Bilateral Amygdala Damage**
**Source: Khalsa et al. (2016)**

Study and result. Cardiorespiratory manipulations elicited panic in individuals with bilateral amygdala damage, indicating that fear sensations are organized through multiple pathways.

GNWT interpretation. GNWT can incorporate subcortical and cortical pathways leading to global access without requiring a single emotional node.

MEM interpretation and discriminating value. MEM predicts a distributed embodied system in which an intense internal signal can activate an affective percept and an action programme despite damage to one structure. The finding argues against simple localizationism, but is compatible with both MEM and multilevel interoceptive theories.

**18. Direct Stimulation and the Intensity of Affect**
**Source: Yih et al. (2019)**

Study and result. Intracranial stimulation at different brain sites produced changes in affective experience, particularly in its intensity, without a simple one-to-one assignment of each emotion to a single centre.

GNWT interpretation. GNWT can interpret stimulation as altering the content or arousal of a globally available state.

MEM interpretation and discriminating value. MEM interprets it as a perturbation of components of a distributed bodily–affective representation that becomes an experienced feeling through re-entry and modifies active associations. Simultaneous measurements of modality-specific reconstruction, access, and report would be most discriminating, because they could establish which stage was causally altered.

**19. Overlapping Maps of Visceral and Emotional Sensations**
**Source: Soulier et al. (2023)**

Study and result. Direct cortical stimulation elicited visceral and emotional sensations in partly overlapping regions, suggesting shared or strongly coupled representations.

GNWT interpretation. Multilevel GNWT can incorporate these representations into globally available contents.

MEM interpretation and discriminating value. In MEM, overlap is expected when interoceptive, affective, and situational components belong to related subsemblions and activation of one reactivates others. The finding supports a distributed architecture, but does not determine whether phenomenality depends on global access, modality-specific re-entry, or both.

**20. Controlled Breathing Practices, Mood, and Arousal**
**Source: Balban et al. (2023)**

Study and result. Brief daily breathing practices altered mood and physiological arousal, with differences among breathing protocols.

GNWT interpretation. GNWT can characterize the effect as a sustained modulation of global state, attention, and content availability.

MEM interpretation and discriminating value. MEM interprets breathing as a physical change in interoceptive input that updates valence, selection parameters, and subsequent perceptual associations.

The data support body–affect coupling. Theoretical adjudication requires a dose–response model and a test of whether changes in modality-specific representation mediate the change in mood.

**21. Do Emotions Have Fixed Autonomic 'Fingerprints'?**
**Source: Siegel et al. (2018)**

Study and result. The meta-analysis did not confirm simple, fixed, and unambiguous autonomic profiles for basic emotion categories; patterns depended on context and method.

GNWT interpretation. GNWT does not require fixed peripheral signatures and can represent emotion as distributed, context-dependent content.

MEM interpretation and discriminating value. MEM should not identify valence with a single physiological profile. An emotional semblion may integrate different interoceptive, situational, and mnemonic configurations depending on the organism's history. The result weakens a strong version of universal somatic signatures but supports context-sensitive variants of MEM and interoceptive predictive processing.

**22. Absence of a Stable Facial-Mimicry Effect**
**Source: Wołoszyn, Hohol, and Winkielman (2026)**

Study and result. The study found no evidence that restricting facial mimicry impaired emotion recognition or altered ratings of the valence and arousal of human affective vocalizations and instrumental sounds.

GNWT interpretation. For GNWT, this is a neutral result: the theory does not require mimicry as a condition of access.

MEM interpretation and discriminating value. For MEM, the result requires a weak, context-sensitive version of coupling. Facial proprioception is one of many possible components and may fail to attain the weight required to change the winning representation. If MEM predicted a necessary and strong effect of mimicry, the data would weaken it. The moderate version remains compatible, but is less risky.

**23. Interoception as a Mechanism of Emotion and Cognition**
**Source: Greenwood and Garfinkel (2025)**

Study and result. The review synthesizes evidence that bodily signals influence perception, emotion, memory, and decision-making, emphasizing context dependence and individual differences.

GNWT interpretation. Contemporary GNWT can incorporate interoception as both content and a modulatory influence on the global network, although this is not a classical focus of the theory.

MEM interpretation and discriminating value. The scope of the review corresponds closely to MEM: bodily state influences representational activation, valence, and action. At the same time, the evidence is equally natural for interoceptive predictive processing. The review organizes a class of evidence but is not a direct test of competing mechanisms.

**24. Interoception and Emotion: Integrative Frameworks**
**Source: Critchley and Garfinkel (2017)**

Study and result. The review describes bodily afferent pathways, cortical interoceptive representations, and their role in feeling and behavioural regulation.

GNWT interpretation. GNWT can explain the access to and use of such representations, but this does not determine the genesis of their modality-specific quality.

MEM interpretation and discriminating value. MEM locates an important component of phenomenality in the perception and reconstruction of internal states and in their association with situation and action. Functional-level convergence is substantial; manipulations separating interoceptive accuracy from global availability are required.

25. **A Roadmap for Research on Interoception and Mental Health**
**Source: Khalsa et al. (2018)**

Study and result. The consensus statement identified interoception as a multilevel construct encompassing bodily signals, their detection, interpretation, awareness, and regulation, and specified the need for methodological standardization.

GNWT interpretation. These distinctions enable GNWT to model which representations become globally available and reportable.

MEM interpretation and discriminating value. MEM can assign the respective levels to receptors, FFS, semblion activation, RP, secondary perception, and regulatory-programme selection. The paper provides a methodological framework rather than adjudicating between theories. It is particularly important for operationalizing theory-specific tests of MEM.

## 8. Structured Results of the Comparison

### 8.1. Profiles of Criteria and Study Groups

Items 13–25 increase the weight assigned in the corpus to evidence concerning interoception, affect, and organismic regulation. This sampling favours models that explicitly formalize these variables, particularly MEM. Mean fit was 5.09 for MEM and 3.00 for GNWT. The narrower stated scope of classical GNWT in this domain primarily indicates a different theoretical emphasis, rather than an empirical challenge to the theory.

Table 3 presents M and K for the sixteen criteria. Small differences in K should not be interpreted as a ranking of truth; the profile across criteria is more informative. GNWT performs best on threshold dynamics, chronometry, electrophysiology, and reporting conditions, whereas MEM has a broader M in the domains of modality, interoception, learning, and action, but lacks theory-specific evidence.

**Table 3. Explanatory scope M and conservative fit K for sixteen criteria. For each theory, cells report M / K.**

| Criterion | E | MEM M/K | GNWT M/K | Highest M |
|---|---|---|---|---|
| P1 | 4.65 | 6 / 4.65 | 5 / 4.65 | MEM |
| P2 | 4.33 | 6 / 4.33 | 3 / 3.00 | MEM |
| P3 | 4.66 | 6 / 4.66 | 5 / 4.66 | MEM |
| P4 | 3.75 | 6 / 3.75 | 4 / 3.75 | MEM |
| A1 | 4.04 | 6 / 4.04 | 6 / 4.04 | tie |
| A2 | 4.36 | 5 / 4.36 | 6 / 4.36 | GNWT |
| A3 | 4.41 | 5 / 4.41 | 6 / 4.41 | GNWT |
| A4 | 4.08 | 6 / 4.08 | 6 / 4.08 | tie |
| S1 | 4.39 | 6 / 4.39 | 6 / 4.39 | tie |
| S2 | 5.14 | 5 / 5.00 | 5 / 5.00 | tie |
| S3 | 3.31 | 6 / 3.31 | 4 / 3.31 | MEM |
| S4 | 4.24 | 6 / 4.24 | 3 / 3.00 | MEM |

| Criterion | E | MEM M/K | GNWT M/K | Highest M |
|---|---|---|---|---|
| B1 | 4.52 | 6 / 4.52 | 6 / 4.52 | tie |
| B2 | 5.04 | 6 / 5.04 | 5 / 5.00 | MEM |
| B3 | 4.67 | 4 / 4.00 | 6 / 4.67 | GNWT |
| B4 | 4.80 | 4 / 4.00 | 5 / 4.80 | GNWT |

*Note. M denotes the theory's stated explanatory scope, E—consistency with the body of evidence, and K = min(M, E)—a conservative fit bounded by the weaker of these two components. The ratings constitute a structured synthesis, not a meta-analysis.*

**Table 4. Mean profiles across four dimensions. For each theory, cells report M / K.**

| Dimension | E | MEM | GNWT |
|---|---|---|---|
| P — phenomenality | 4.35 | 6.00 / 4.35 | 4.25 / 4.02 |
| A — access | 4.22 | 5.50 / 4.22 | 6.00 / 4.22 |
| S — efficacy | 4.27 | 5.75 / 4.23 | 4.50 / 3.92 |
| B — biology | 4.76 | 5.00 / 4.39 | 5.50 / 4.75 |

*Note. Each theory cell reports mean M / K. M denotes stated explanatory scope, whereas K = min(M, E) is the conservative fit.*

## 9. Discussion

### 9.1. Can the Classical GNWT Findings Be Explained by MEM?

Yes—at the level of qualitative mechanistic compatibility, most findings can be explained by WTA selection, threshold stabilization of a semblion, recurrent processes, associative propagation, and linkage to a reporting programme. The data therefore support a broader class of recurrent–selective architectures rather than GNWT alone. This does not imply equal precision: GNWT retains an advantage in chronometry, operationalization of ignition, and decades of task-specific model fitting.

### 9.2. Access Consciousness

MEM contains candidate counterparts of GNWT's access-related functions: a dominant semblion can be selected, maintained, associated with memory and language, and used to guide action. It has not, however, been demonstrated that this implementation provides measurable globality, capacity, robustness, and flexible routing at the level achieved by GNWT models. Similar functional scope therefore does not imply comparable empirical maturity.

### 9.3. Qualia and Phenomenal Consciousness

MEM's distinctive ambition concerns the genesis of experiential quality. A semblion is proposed to remain grounded in modality-specific receptor traces, while re-entry reinstates sensory and interoceptive configurations. GNWT primarily explains when such content becomes available and useful. A demonstrated necessity of modality-specific reconstruction despite preserved access would support MEM; preservation of experience despite blocking such reconstruction and preventing secondary perception would weaken the strong version of the model.

### 9.4. Interoception, Allostasis, and Motivation

Explicit allostatic, affective, and motivational variables broaden MEM relative to classical GNWT, but do not confer uniqueness relative to active inference. In interoceptive predictive processing, bodily states are objects of hierarchical inference, and action can realize preferred states by minimizing expected free energy. In active inference, preferences, policies, and epistemic value extend the model beyond exteroception, while recent accounts develop it as a theory of sentient behaviour (Pezzulo et al., 2024). MEM's potential advantage may therefore lie not in acknowledging the body as such, but in the specific combination of ordinary FFS neuronal discharges, semblion-based memory, procedural branches, and modality-specific reconstruction. It is this package, rather than the general notion of embodiment, that requires testing.

### 9.5. Image Context and Image Valence in MEM

The context of an image may be carried by co-active subsemblions representing the scene, episodic memory, body position, the current task, and the preceding sequence, linked laterally and recurrently to the modality-specific representation. Image valence may be implemented by associated interoceptive and affective traces, modulatory signals related to allostatic violations, and consolidated connections to avoidance, approach, or regulatory programmes. There is no need to postulate a separate symbol for 'context' or 'value': these are properties of the state of the entire active configuration and of its history of plasticity. The hypothesis becomes testable when it specifies particular populations, delays, flow directions, and necessity conditions.

### 9.6. What Does the Present 'Advantage' of MEM Mean?

The present evidence supports three moderate conclusions. First, MEM has an architecture sufficiently rich to constitute a serious candidate model of access consciousness. Second, the reinterpretation shows that many canonical GNWT findings are not theoretically exclusive. Third, MEM proposes mechanisms for modality, affect, and phenomenality that extend beyond the classical GNWT core. This does not yet justify the claim that MEM is better supported. The most accurate formulation is a hypothesis of explanatory advantage that requires quantitative model comparison and direct tests of distinctive predictions.

## 10. A Programme of Experiments to Discriminate between MEM and GNWT

Analysis of the findings indicates that, within the selected corpus, the two theories achieve comparable results. Demonstrating an advantage for the newer theory requires a series of studies targeting the distinctive features of both theories, particularly their capacity to explain phenomenal consciousness. Discriminating studies should manipulate processes that the theories regard as constitutive, rather than merely measuring their shared correlates. Each project requires preregistration of the theory version, specification of parameters before analysis, no-report measures supplemented by delayed report, and control of attention, confidence, working memory, and behavioural responses.

### 10.1. Dissociating Modality-Specific Re-entry from Global Availability

In a masking or binocular-rivalry task, feedback connections to early sensory fields and long-range frontoparietal connections should be attenuated independently using temporally precise TMS, layer-specific stimulation, or closed-loop electrophysiological intervention. MEM predicts a qualitatively specific impoverishment or alteration of content after disruption of modality-specific RP even when task access is preserved. Strong GNWT predicts that preserved global access should sustain conscious content, whereas disruption of broadcasting should abolish flexible use regardless of local recurrence.

### 10.2. Causal Order during Spontaneous Changes in Perceptual Content

In a paradigm without stimulus change, such as binocular rivalry, the direction and delays of information flow among early modality-specific cortex, high-level representations, the thalamus, and prefrontal cortex should be estimated. Analyses must be combined with intervention, because statistical directionality alone does not establish causation. MEM predicts the cycle: a change in semblion competition, modality-specific reconstruction, secondary perception, and only then report. GNWT predicts nonlinear entry into the global network, after which the content is stabilized and broadcast.

### 10.3. Interoceptive Gating of Quality with Preserved Performance

Near-threshold stimuli should be combined with controlled changes in cardiac phase, breathing, or visceral afferent signals. Matching task difficulty and response criterion is crucial. MEM predicts a change in perceptual valence, vividness, or quality proportional to activation of associated interoceptive subsemblions. GNWT may predict a change in the probability of access, but not necessarily a modality-specific change in content; interoceptive predictive processing predicts effects dependent on precision and prior beliefs. Competing models should predict complete response distributions.

### 10.4. Reconstruction of Endogenous Content without Current Input

During imagery, recall, dreaming, or hallucination, content structure should be decoded in high-level and early modality-specific fields while each level is selectively disrupted. MEM predicts that preventing the appropriate topographic reconstruction should alter a specific quality of the image, although some semantic access may remain. GNWT primarily predicts dependence of reportability and stability on global availability. The test should distinguish content accuracy, vividness, confidence, and availability for action.

### 10.5. Geometry of an Active Semblion

During the learning of novel arbitrary objects, the formation of representations should be traced from feature fields, through selective neurons and cross-modal associations, to action programmes and affective signals. MEM predicts partly overlapping multilayer traces in which a small apical population indexes a distributed pattern and can reinstate it recurrently. GNWT does not require this representational geometry. Causal reactivation of the apex should reinstate specific modality-related, contextual, and procedural components; failure of such reconstruction would weaken the strong semblion hypothesis.

### 10.6. Globality versus Rich Associativity

To distinguish broadcasting from propagation through overlapping associations, tasks should be constructed with novel recipients of information that did not participate in learning the percept. GNWT predicts flexible routing of the winning content to multiple unprepared task systems. MEM must explain whether and how a new recipient gains access without a pre-existing semblion connection. Rapid arbitrary reconfiguration would favour a workspace mechanism; dependence on associative history would favour the MEM mechanism.

### 10.7. Double Dissociation of Quality and Report

Investigators should seek one intervention that alters modality-specific quality while preserving identification and action, and another that abolishes report or flexible use while preserving markers of modality-specific experience. Delayed report, autonomic reflexes, perceptual aftereffects, and incidental learning should be analysed jointly rather than treated as individual proxies for consciousness. A double dissociation would provide stronger evidence than additional correlations involving P3b or connectivity.

### 10.8. Quantitative Model Comparison

GNWT and MEM models should receive common inputs, identical training data, and an observation set comprising behaviour, EEG/MEG, fMRI, autonomic signals, and reports of experiential quality. Parameters should be estimated on a subset of tasks and used to predict other tasks and interventions. Information criteria, cross-validation, and parameter-identifiability analyses should penalize flexible post hoc fitting. Only an out-of-sample advantage replicated across laboratories would warrant a claim of greater explanatory power.

### 10.9. Criteria for Falsifying MEM

Strong MEM would be weakened if: (a) conscious, modality-rich experience persisted after effective disruption of the postulated re-entry; (b) activation of a candidate semblion failed to reinstate associated sensory and procedural components; (c) interoceptive changes affected only decision or report and never quality, valence, or learning; or (d) global availability explained all effects without additional predictive value from MEM variables. Moderate versions of the theory must specify in advance which of these conditions they regard as central, thereby avoiding post-result shifts in the target claim.

## 11. Conclusions

GNWT remains the most empirically mature of the programmes of conscious access compared here. Its canonical predictions concern selection, a nonlinear threshold, maintenance, long-range communication, and flexible use of information. A decades-long body of dedicated research confirms the existence of these phenomena, although their interpretation as conditions sufficient for phenomenal experience remains controversial.

MEM proposes functional counterparts of the principal GNWT operations: WTA competition, a working-memory-dominant semblion, FFS–RP cycles, associative propagation, and linkage to language and action programmes. In this sense, its explanatory scope for access consciousness need not be narrower. Not all of these counterparts are equally developed quantitatively, however; measurable globality, the ignition threshold, working-memory gating, and arbitrary routing particularly require formalization and empirical testing.

Analysis of GNWT research shows that its findings are also substantially compatible with MEM. This argues against the exclusivity of a single interpretation and supports a common class of recurrent–selective mechanisms. It is not yet theory-specific evidence for MEM, because post hoc compatibility cannot substitute for prospective prediction.

MEM’s greatest additional potential concerns the receptor-grounded structure of semblions, modality-specific re-entry, secondary perception, and the coupling of content with interoception, valence, allostasis, and action programmes. This combination may provide an explanation of qualia and phenomenal consciousness broader than the classical GNWT core.

The most justified conclusion, therefore, is not that the analysis demonstrates the superiority of MEM. Rather, it supports a hypothesis of explanatory advantage that requires quantitative testing. MEM’s recognition as an empirically mature theory on an equal footing will depend on a body of preregistered studies dissociating modality-specific reconstruction, global access, report, and interoceptive valuation, and on demonstrating that a single parameterized model predicts the results of multiple tasks and interventions better than its competitors.

## Appendix A. Publication Rating Matrices

The matrices preserve the order of the 25 publications and the weights adopted in the section. The values assigned are analytical ratings, not effect estimators. Row E reports the weighted mean correspondence of the corpus with the criteria.

The consolidated matrix permits verification of the ratings adopted in the analysis and critical assessment of the conclusions derived from them.

**Table 5. Ratings of publications 1–25 across the dimensions of phenomenal consciousness (P1–P4), access consciousness (A1–A4), system efficacy (S1–S4), and biological plausibility (B1–B4).**

| Study | | Phenomenal consciousness | | | | Access consciousness | | | | System efficacy | | | | Biological plausibility | | | |
|---|---|---|---|---|---|---|---|---|---|---|---|---|---|---|---|---|---|
| **No.** | **Weight** | **P1** | **P2** | **P3** | **P4** | **A1** | **A2** | **A3** | **A4** | **S1** | **S2** | **S3** | **S4** | **B1** | **B2** | **B3** | **B4** |
| **1** | 1 | 5 | 3 | 4 | 3 | 5 | 6 | 3 | 3 | 4 | 5 | 3 | 3 | 3 | 3 | 3 | 3 |
| **2** | 1 | 6 | 3 | 5 | 3 | 5 | 5 | 5 | 4 | 6 | 5 | 3 | 3 | 5 | 5 | 6 | 3 |
| **3** | 1 | 6 | 3 | 6 | 3 | 4 | 5 | 6 | 5 | 6 | 4 | 3 | 3 | 6 | 5 | 6 | 3 |
| **4** | 1 | 5 | 4 | 5 | 4 | 4 | 5 | 6 | 5 | 6 | 6 | 5 | 3 | 5 | 6 | 6 | 4 |
| **5** | 1 | 5 | 3 | 4 | 5 | 4 | 5 | 5 | 5 | 4 | 5 | 3 | 3 | 5 | 5 | 5 | 6 |
| **6** | 1 | 4 | 4 | 6 | 6 | 6 | 4 | 5 | 5 | 4 | 5 | 3 | 4 | 4 | 5 | 4 | 6 |
| **7** | 1 | 5 | 3 | 5 | 4 | 5 | 6 | 6 | 6 | 6 | 6 | 3 | 4 | 6 | 5 | 6 | 5 |
| **8** | 1 | 5 | 4 | 6 | 6 | 4 | 4 | 6 | 5 | 5 | 5 | 3 | 5 | 5 | 6 | 6 | 6 |
| **9** | 0.75 | 4 | 3 | 6 | 3 | 5 | 6 | 6 | 4 | 4 | 5 | 3 | 3 | 6 | 4 | 5 | 4 |
| **10** | 0.5 | 4 | 3 | 6 | 3 | 6 | 6 | 6 | 5 | 5 | 6 | 4 | 4 | 6 | 5 | 5 | 3 |
| **11** | 0.5 | 5 | 3 | 6 | 3 | 6 | 6 | 6 | 6 | 6 | 5 | 4 | 3 | 6 | 5 | 6 | 3 |
| **12** | 0.25 | 6 | 4 | 6 | 6 | 4 | 4 | 4 | 5 | 6 | 4 | 3 | 3 | 6 | 4 | 5 | 5 |
| **13** | 1 | 5 | 6 | 4 | 4 | 4 | 4 | 4 | 4 | 4 | 6 | 3 | 6 | 5 | 6 | 5 | 6 |
| **14** | 1 | 4 | 6 | 4 | 3 | 3 | 3 | 3 | 3 | 4 | 5 | 3 | 4 | 3 | 4 | 3 | 5 |
| **15** | 1 | 4 | 5 | 4 | 3 | 3 | 3 | 3 | 3 | 4 | 5 | 3 | 4 | 3 | 4 | 3 | 5 |
| **16** | 1 | 5 | 6 | 4 | 3 | 5 | 5 | 4 | 4 | 5 | 6 | 3 | 5 | 5 | 6 | 6 | 4 |
| **17** | 1 | 5 | 6 | 4 | 4 | 3 | 4 | 4 | 4 | 4 | 5 | 3 | 6 | 4 | 6 | 4 | 6 |
| **18** | 1 | 5 | 6 | 5 | 4 | 3 | 4 | 4 | 3 | 3 | 4 | 3 | 5 | 5 | 6 | 5 | 6 |
| **19** | 1 | 5 | 6 | 5 | 4 | 3 | 4 | 4 | 3 | 3 | 4 | 3 | 5 | 5 | 6 | 5 | 6 |
| **20** | 1 | 4 | 5 | 4 | 3 | 3 | 3 | 3 | 4 | 4 | 5 | 4 | 6 | 3 | 5 | 4 | 5 |
| **21** | 1 | 3 | 4 | 3 | 3 | 3 | 3 | 3 | 3 | 3 | 6 | 4 | 5 | 3 | 5 | 3 | 5 |
| **22** | 1 | 3 | 2 | 3 | 3 | 3 | 3 | 3 | 3 | 3 | 5 | 3 | 3 | 3 | 3 | 3 | 5 |
| **23** | 0.25 | 5 | 6 | 5 | 4 | 4 | 4 | 4 | 4 | 4 | 6 | 5 | 6 | 5 | 6 | 5 | 4 |
| **24** | 0.25 | 5 | 6 | 5 | 4 | 4 | 4 | 4 | 4 | 4 | 6 | 5 | 6 | 5 | 6 | 5 | 4 |
| **25** | 0.25 | 4 | 6 | 5 | 4 | 4 | 4 | 4 | 4 | 4 | 6 | 5 | 6 | 5 | 6 | 4 | 5 |
| **E** | — | **4.65** | **4.33** | **4.66** | **3.75** | **4.04** | **4.36** | **4.41** | **4.08** | **4.39** | **5.14** | **3.31** | **4.24** | **4.52** | **5.04** | **4.67** | **4.80** |

*Note. Studies were rated on a scale from 1 to 6. Publication weights: empirical study or meta-analysis—1; hybrid study—0.75; computational model—0.50; theoretical paper or review—0.25. E denotes the weighted mean.*